\documentstyle[twocolumn,prl,aps]{revtex}
\begin{document}
\draft 
\title{Quantum gambling using two nonorthogonal states
 }
\author{WonYoung Hwang \cite{email},
 Doyeol (David) Ahn \cite{byline},
 and Sung Woo Hwang \cite{byline2}}

\address{ Institute of Quantum Information Processing
and Systems,
 University of Seoul 90, Jeonnong, Tongdaemoon,
 Seoul 130-743, Korea
}
\maketitle

\begin{abstract}
We give a (remote) quantum gambling scheme that makes use of
the fact that quantum nonorthogonal states cannot be
distinguished with certainty.
In the proposed scheme, two participants Alice and Bob
can be regarded as playing a game of making guesses
on identities of quantum states that are in one of two
given nonorthogonal states:
if Bob makes a correct (an incorrect)
guess on the identity
of a quantum state that Alice has sent, he wins (loses).
It is shown that the proposed scheme is secure
against the nonentanglement attack. It can also be shown
heuristically that the scheme is secure in the case of
the entanglement attack.
\end{abstract}


\narrowtext

A fundamental property of quantum bits (qubits) that
differs from those of classical bits is that unknown
qubits cannot be copied with unit efficiency
\cite{woot,diek} (the no-cloning theorem).
Another related property of qubits is 
that nonorthogonal qubits cannot be
distinguished with certainty \cite{levi,hole,ivan,pere,hutt,mass}.
The no-cloning theorem is
the basis for the success of the Bennett-Brassard 1984
quantum-key-distribution scheme \cite{ben2}.
Therefore, it is interesting to search for
quantum protocols utilizing  the property that
nonorthogonal qubits cannot be distinguished with
certainty.
Bennett's other quantum-key-distribution scheme
\cite{ben3} indeed utilizes this property.
On the other hand, a (remote) 
quantum gambling scheme  has been
found by Goldenberg {\it et al}. recently \cite{gold}.

In this paper we propose another (remote) quantum gambling
scheme that makes use of the fact that two nonorthogonal
qubits cannot be distinguished with certainty.
In the proposed scheme, two participants Alice and Bob
can be regarded as playing a game of making guesses
on identities of quantum states that are in one of two
given nonorthogonal states;
Alice randomly sends one of
two nonorthogonal qubits, say, $|0\rangle$ and
$|\bar{0}\rangle$. [In this paper, 
$|\bar{0}\rangle \equiv (1/\sqrt{2})(|0\rangle+|1\rangle)$ and 
$|\bar{1}\rangle \equiv (1/\sqrt{2})(|0\rangle-|1\rangle).$] 
If Bob makes a correct guess,
he wins. If not, he loses.
Due to the fact that two nonorthogonal qubits cannot be
distinguished with certainty, it is easy to see that
there is no way for Bob to cheat.
Alice might try to increase her gain
by sending some qubits
other than $|0\rangle$ and $|\bar{0}\rangle$.
There are two kinds of attacks.
In nonentanglement attacks, qubits sent to Bob are
not entangled with Alice's. In entanglement attacks
\cite{lo2,may2} (or EPR attack), qubits sent to Bob are
highly entangled with hers. We show that the scheme is
secure in the case of nonentanglement attacks. In the case
of entanglement attacks, however, we heuristically show the
security of the scheme. It is true that a quantum
cryptographic scheme is of little use without
security proof against
all attacks including entanglement attacks. And what
makes it complicated to prove security of a
quantum cryptographic scheme is
the entanglement attack \cite{lo2,may2}.
Our security proof of the scheme against the
entanglement attack is heuristic. However, since the idea
behind the proof is simple,
we believe that a rigorous one will be found, like in the
case of the quantum key distributions 
\cite{maye,lo,biha,sho2}.

The difference between our scheme and
the original one \cite{gold} is that the former relies on
the fact that two nonorthogonal states cannot be distinguished with
certainty while the latter relies on gerneral quantum 
mechanical laws.d
Another difference is that no quantum system
needs to be additionally
sent in checking steps in our scheme while it needs to be
in the original scheme.

Now, let us  describe the scheme more precisely.\\
(1) Alice randomly chooses one between two
nonorthogonal qubits $|0\rangle$ and $|\bar{0}\rangle$,
and sends it to Bob
\cite{fnt1}.\\
(2) On the qubit he receives, Bob performs a
measurement by which he can obtain maximal
probability $p$ of correctly guessing the identity of
the qubit.\\
(3) On basis of the measurement's results, he
makes a guess on which one the qubit is
and announces it to Alice.\\
(4) If he made a correct (an incorrect) guess, Alice
announces he has won (lost). \\
(5) When Bob has won, Alice gives
him one coin.  When he has lost,
Bob gives her $p/(1-p)$ coins. \\
However, after the first step,
Bob follows the following ones instead of steps $(2)-(5)$,
at randomly chosen instances with a rate $r$
($0<r \ll 1$).\\
(These checking steps are similar to those of the 
original work of Goldenberg {\it et al.} \cite{gold}.)\\
($2^{\prime}$) Bob performs no measurement on the qubit and stores
it.\\
($3^{\prime}$) He announces his randomly chosen guess on identity
of the qubit.\\
($4^{\prime}$) Do the same thing as step (4).\\
($5^{\prime}$) In the previous step,
Alice has actually revealed which one
she chose to tell him the qubit is
(regardless of her honesty).
When it is $|0\rangle$,
Bob performs $\hat{S}_z$
($\hat{S}_z$ an orthogonal measurement
that composes of two projection operators
$|0\rangle \langle0|$ and $|1\rangle \langle 1|$ or
$\{|0 \rangle \langle 0|,|1 \rangle \langle 1|\}$.)
If the outcome is $|1\rangle$, Bob announces that he
performed $\hat{S}_z$ and got $|1\rangle$
as an outcome. Then Alice must give him $R$
($\gg 1 $) coins.
If the outcome is $|0\rangle$, Bob says nothing about
which measurement he performed and follows step (5).
In the case of $|\bar{0}\rangle$,
similar things are done with
$\hat{S}_x$
($=\{|\bar{0}\rangle \langle\bar{0}|,
     |\bar{1}\rangle \langle\bar{1}|\}$).\\
In step (2), it is important for Bob to perform
the optimal measurement that assures maximal probability
$p$ of correctly guessing the identity of
the qubit in order to assure
his maximal gain.
Although it is known that
average information gain is constrained by the
Levitin-Holevo bound \cite{levi,hole,per2}, to find the
optimal one is not an
easy task. Fortunately, however,
in the case of two nonorthogonal
qubits, the measurement giving maximal information gain
is well known \cite{hutt,mass}:
a measurement $\{|\tilde{0}\rangle \langle\tilde{0}|,
|\tilde{1}\rangle \langle\tilde{1}|\}$,
where
$|\tilde{0}\rangle$ and $|\tilde{1}\rangle$
are qubits corresponding to a vector
$(1/\sqrt{2})(\hat{z}- \hat{x})$
and $(1/\sqrt{2})(-\hat{z}+ \hat{x})$, respectively,
in the Bloch sphere representation,
where a single qubit density operator
$\rho_B= (1/2)({\bf 1}+ \hat{r} \cdot {\bf \vec{\sigma}})$.
Here ${\bf 1}$  is the identity operator,
$\hat{r}$ is a Bloch vector,
${\bf \vec{\sigma}}= (\sigma_x, \sigma_y, \sigma_z)$,
and $\sigma_x, \sigma_y, \sigma_z$ are the Pauli operators.
(See Fig. 1 of Ref. \cite{hutt} or Fig. 2 of
Ref. \cite{mass}.)
Since information gain is maximal
if and only if $p$ is maximal, a measurement with maximal
information gain is what maximizes $p$.
Thus the measurement
$\{|\tilde{0}\rangle \langle\tilde{0}|,
|\tilde{1}\rangle \langle\tilde{1}|\}$
is the optimal one.
For maximal $p$, Bob does as the following.
When the outcome is
$|\tilde{0}\rangle$ ($|\tilde{1}\rangle$), he makes
a guess that the qubit is
$|0\rangle$ ($|\bar{0}\rangle$).
Then the probability $p$ of
correctly guessing the qubit is given by
$p= | \langle\tilde{0}|0\rangle |^2
 = | \langle\tilde{1}|1\rangle |^2
 = \cos^2 (\pi/8)$.

Now let us show how each player's average gain is assured.
First it is clear by definition that
Bob can do nothing better than
performing the optimal measurement,
as long as Alice prepares the specified qubits.
In the scheme, the numbers
of coins that Alice and Bob pay
are adjusted so that no one gains when Bob's
win probability is $p$.
Thus Bob's gain $G_B$ cannot be greater
than zero, that is, $G_B \leq 0$.
Next let us consider Alice's strategy.
As noted above, we first show the security against
nonentanglement attacks.
In the most general nonentanglement attacks,
Alice randomly generates each qubit in state $|i\rangle$
with a probability $p_i$.
Here $|i\rangle$'s are arbitrarily specified states
of qubits,
$i=1,2,...,N$ and $\sum_i^N p_i=1$. However, since Bob has
no knowledge about which $|i\rangle$ Alice selected at each
instance, his treatments on qubits become equal for all
qubits. Thus it is
sufficient to show the security for a qubit in an
arbitrary state
$|j\rangle= a|0\rangle + b|1\rangle$.
($a$ and $b$ are some complex numbers with a constraint
$|a|^2+ |b|^2=1$.)
First we do it for states within the $z$-$x$ plane,
$|j\rangle= \cos (\theta/2) |0\rangle +
            \sin (\theta/2) |1\rangle$.
Later we will generalize the argument to the former
case.
Let us consider the following. In steps
$(2^{\prime})-(5^{\prime})$, 
Bob checks at randomly chosen instances whether
Alice has really sent
$|0\rangle$ or $|\bar{0}\rangle$ by performing measurements
$\hat{S}_z$ or $\hat{S}_z$, respectively. If the
measurement's outcomes are $|0\rangle$ or $|\bar{0}\rangle$
($|1\rangle$ or $|\bar{1}\rangle$), Alice has passed
(not passed) the test. When not passed, Alice must give
him $R (\gg 1)$ coins \cite{fnt2}.

Roughly speaking,
Alice can do nothing but preparing
either $|0\rangle$ or $|\bar{0}\rangle$ and honestly
tell the identity of the state to him later.
Otherwise she sometimes must pay $R$ coins to him,
decreasing her gain.
Let us consider this point more precisely.
First we estimate the upper bound of Alice's gain $G_A$.
It is clear that
\begin{equation}
\label{b}
G_A \leq \mbox{max} \{G_A^{0}, G_A^{\bar{0}} \},
\end{equation}
where
max$ \{x,y\}$ denotes the maximal one between $x$ and
$y$, and  $G_A^{0}$ ($G_A^{\bar{0}}$) is Alice's gain when
she insists that $|j\rangle$ is
$|0\rangle$ ($|\bar{0}\rangle$).
When he has performed the measurements already,
he has no way of detecting Alice's cheating.
So, Alice's maximal gain in this
case is $p/(1-p)$.
However, it is clear that $G_A$ is bounded by
$p/(1-p)$ in any case. When he has preserved the qubits
following the checking steps, Alice's cheating can be
statistically detected. This case
contributes to Alice's gain
by a largely negative term whose modulus is proportional
to the product
of the rate $r$ of checking steps, the probability that
$|1\rangle$ or $|\bar{1}\rangle$
is detected, and the number of coins $R$
she must pay when it is detected, namely
$-r|\langle 1|j\rangle|^2R$ or
$-r|\langle \bar{1}|j\rangle|^2R$.
However, here we should take into account the fact that
Alice obtains partial information about whether Bob has
performed the measurement:
let $f_u$ be Alice's estimation of the probability that
Bob did not perform the measurement.
With no information, $f_u $ is $r$.
However, Bob's announced guess gives her
partial information on his measurement's result
if he did. This information can be used to
make a better estimate of $f_u$. For example,
in the case where Alice sends $|j\rangle$
and Bob performs the optimal measurement
$\{|\tilde{0}\rangle \langle\tilde{0}|,
|\tilde{1}\rangle \langle\tilde{1}|\}$,
we obtain using the Bayes's rule
that $f_u= (r/2)/[(r/2)
+(1-r)|\langle \tilde{0}|j\rangle|^2]$
when his guess is $|0\rangle$.
However, it is clear that $f_u \geq r/2$:
when Bob did not perform the measurement,
he simply guesses it with equal probabilities
regardless of what he received.
Thus by the Bayes's rule, Alice can see that there remains 
a probability greater than $r/2$ that Bob did not perform
the measurement. The relation $f_u \geq r/2$ also holds
for the entanglement attacks, since it is satisfied for any
$|j\rangle$ as shown above (refer to the related discussion 
on entanglement-attack later). Combining above facts, we 
obtain
\begin{equation}
\label{c}
G_A^{0} \leq \frac{p}{1-p}
     -\frac{r}{2}|\langle 1|j\rangle|^2 R
\end{equation}
and
\begin{equation}
\label{d}
G_A^{\bar{0}} \leq \frac{p}{1-p}
     -\frac{r}{2} |\langle \bar{1}|j\rangle|^2 R.
\end{equation}
From Eqs. (\ref{b})-(\ref{d}), we can see, 
in order that $G_A$ be non-negative the following two
conditions must be satisfied.
(1) Either $|\langle 1|j\rangle|^2 \sim (1/rR) \ll 1$
(that is, $|j\rangle \sim |0\rangle$)
or $|\langle \bar{1}|j\rangle|^2 \sim (1/rR) \ll 1$
(that is, $|j\rangle \sim |\bar{0}\rangle$);
(2) Between $|0\rangle$ and $|\bar{0}\rangle$,
Alice chooses what is
nearer to $|j\rangle$. Then she pretends in the step (4)'s
that it is the qubit sent to Bob.
Otherwise, $G_A$ will be dominated by the
negative second term in the right-hand sides of Eqs.
(\ref{c}) and (\ref{d}).

Alice might increase her gain by
sending a qubit that slightly differs from either
$|0 \rangle$ or $|\bar{0} \rangle$. However, the gain
can be made negligible by making $R$ large, as we show
in the following.
Let us consider the case where Alice prepares $|j\rangle$
($\sim |0\rangle$)
and later tells him that it is $|0\rangle$, for example.
In this case, $G_A$ for a given $r$ and $R$ is given by
\begin{eqnarray}
\label{e}
G_A &=&
(1-r)
\{|\langle\tilde{0}|j \rangle |^2 (-1)
 +|\langle\tilde{1}|j \rangle |^2 \frac{p}{1-p}\}
\nonumber\\
&&-r|\langle 1|j\rangle|^2 R
  +r|\langle 0|j\rangle|^2
 \{ \frac{1}{2} (-1)+
    \frac{1}{2} \frac{p}{1-p}\}
\nonumber\\
&<&
(1-r)
\{[\cos (\frac{\pi}{8} +\frac{\theta}{2})]^2 (-1)
 +[\sin (\frac{\pi}{8} +\frac{\theta}{2})]^2 
\frac{p}{1-p}\}
\nonumber\\
&&-r (\sin \frac{\theta}{2})^2 R + 3r .
\end{eqnarray}
Here the first (second and third) term
in the right-hand sides is due
to normal steps $(1)-(5)$ [checking steps 
$(2^{\prime})-(5^{\prime})$].
We can check Eq. (\ref{e}) by verifying that
$G_A< 3r \sim 0$  when $|j\rangle$ equals $|0\rangle$.
By the two conditions, we might only consider
the case where
$\theta \sim 0$, and thus
we can neglect higher-order terms in Eq. (\ref{e}),
\begin{equation}
\label{f}
G_A <
\frac{\alpha}{1-p} \theta
-\frac{rR}{4} \theta^2+ 3r,
\end{equation}
where $\alpha=\cos(\pi/8)\sin(\pi/8)$
and a small term of order $\theta^2$ is also neglected.
Alice would maximize her gain for given $r$ and $R$.
Maximal value of $G_A$
is obtained when
$\theta= (2\alpha/[1-p])(1/rR)$.
\begin{equation}
\label{g}
G_A^{max}= \frac{\alpha^2}{(1-p)^2} \frac{1}{rR}+ 3r.
\end{equation}
Bob would minimize $G_A^{max}$ for a given $R$.
$G_A^{max}$ has its minimal value
$2\sqrt{3}\alpha/[(1-p)\sqrt{R}]$
when $r=\alpha/[(1-p)\sqrt{3R}]$.
Therefore, if Bob chooses $r= \alpha/[(1-p)\sqrt{3R}]$
then $G_A$ is bounded by the positive term
$2\sqrt{3}\alpha/[(1-p)\sqrt{R}]
\propto 1/\sqrt{R}$
that approaches to zero as $R$ become large,
similarly to the case of the scheme of Goldenberg {\it et al}.
\cite{gold}.

Now we argue that using a qubit outside the
$z$-$x$ plane does not increase Alice's gain:
we can see in  Eq. (\ref{e}) that $G_A$ can only
be increased by making the ratio
$|\langle\tilde{1}|j \rangle |^2/
|\langle\tilde{0}|j \rangle |^2$ large
while keeping $|\langle 1|j \rangle |^2$ a very
small constant.
Let us consider some set of $|j\rangle$'s (not confined
in the $z$-$x$ plane) that give the same value of
$|\langle 1|j \rangle |^2$.
Bloch vectors of this set make a circle around
that of $|1\rangle$. We can see by inspection that
what gives the maximal value of the ratio
$|\langle\tilde{1}|j \rangle |^2/
|\langle\tilde{0}|j \rangle |^2$ lies within the
$z$-$x$ plane.

Now, let us heuristically argue that
the entanglement attacks \cite{lo2,may2} do not work
in the proposed scheme.
Let us consider the case where
Alice prepares pairs of qubits in an entangled state,
\begin{equation}
\label{h}
|\psi\rangle= \frac{1}{\sqrt{2}}
(|0\rangle_A |0\rangle_B + |1\rangle_A |\bar{0}\rangle_B),
\end{equation}
where $A$ and $B$ denote Alice and Bob, respectively.
Alice sends qubits with label $B$ to Bob while storing
those with label $A$. If she performs $\hat{S}_z$,
Bob is given a mixture of $|0\rangle$ and
$|\bar{0}\rangle$ with equal frequency.
Thus if Alice always performs $\hat{S}_z$, the attack
reduces to a nonentanglement attack where she
randomly sends either $|0\rangle$ or $|\bar{0}\rangle$.
Let us consider an example illustrating how performing
measurements much different from $\hat{S}_z$ is not of
benefit for Alice; we can rewrite Eq. (\ref{h}) as
\begin{equation}
\label{i}
|\psi\rangle=
\frac{\sqrt{2+\sqrt{2}}}{2}
|\bar{0}\rangle_A |\alpha\rangle_B+
\frac{\sqrt{2-\sqrt{2}}}{2}
|\bar{1}\rangle_A |\beta\rangle_B,
\end{equation}
where $|\alpha\rangle$ and $|\beta\rangle$
are normalized ones of
$(|0\rangle+ |\bar{0}\rangle)$ and
$(|0\rangle- |\bar{0}\rangle)$, respectively.
Thus if Alice performs $\hat{S}_x$,
either $|\alpha\rangle$ or $|\beta\rangle$
is generated at Bob's site
with probabilities given by Eq. (\ref{i}).
However, since all of $|\langle1|\alpha\rangle|^2$,
$|\langle\bar{1}|\alpha\rangle|^2$,
$|\langle1|\beta\rangle|^2$, and
$|\langle\bar{1}|\beta\rangle|^2$ are of order of one,
$G_A$ becomes much
negative in any case. So Alice would not perform
$\hat{S}_x$. In fact, if Alice is able to change the qubits
between $|0\rangle$ and $|\bar{0}\rangle$ as she likes,
her cheating will always be successful. However, she is not
allowed to do so, since
$|0\rangle \langle0| \neq |\bar{0}\rangle \langle\bar{0}|$
and Bob's reduced density operator
$\rho_B(=\mbox{Tr}_A[\rho_{AB}])$
cannot be changed even with entanglement attacks.

By appropriately choosing her measurement, Alice
can generate at Bob's site any
$\{p_i, |i\rangle \langle i| \}$ satisfying
$\sum_i p_i |i\rangle \langle i|= \rho_B$, where
$\{p_i, |i\rangle \langle i| \}$
denotes a mixture of pure states $|i\rangle \langle i|$
with relative frequency $p_i$
(the theorem of Houghston, Jozsa, and Wootters) \cite{hugh}.
Let
$\rho_B= (1/2)({\bf 1}+ \hat{r} \cdot {\bf \vec{\sigma}})$.
Since $\rho_B= \sum_i p_i |i\rangle \langle i|$ and
$|i\rangle \langle i|=
(1/2)({\bf 1}+ \hat{r}_i \cdot {\bf \vec{\sigma}})$ where
$\hat{r}_i$ is the corresponding Bloch vector, we have
$(1/2)({\bf 1}+ \hat{r} \cdot {\bf \vec{\sigma}})=
(1/2)({\bf 1}+ [\sum_i p_i\hat{r}_i]
\cdot {\bf \vec{\sigma}})$ and thus
\begin{equation}
\label{j}
\hat{r}= \sum_i p_i\hat{r}_i.
\end{equation}
Therefore, for a given $\rho_B$
whose Bloch vector is $\hat{r}$,
Alice can prepare at Bob's site any
mixture $\{p_i, |i\rangle \langle i| \}$
as long as its Bloch
vectors $\hat{r}_i$ satisfy the Eq. (\ref{j}).
However, if Alice always performs a given measurement,
the entanglement attacks reduce to the
nonentanglement attacks: outcomes of measurements on
entangled pairs do not depend on temporal
order of the two participants' measurements. So
we can confine ourselves to the case where Alice measures
first. Then the attack reduces to a nonentanglement
attack where Alice generates
$|i\rangle$ with probability $p_i$.
Alice can only utilize the
entanglement by choosing her measurements according to
Bob's announced guesses. However,
the checking steps also prevent Alice from increasing her
gain: she must choose the measurement that gives some mixture
$\{p_i, |i\rangle \langle i| \}$ at Bob's site where each
$\hat{r}_i$ is nearly the same as either $z$ or $x$.
Otherwise $G_A$ becomes dominated by a much negative
term involving $rR$. Therefore, Alice's freedom in the
choice of measurements is negligible and thus she can
increase her gain by negligible amounts even with the
entanglement attacks.

Although the proposed scheme can be implemented with
currently
available technologies, it is very sensitive to errors.
So before methods for reducing decoherence,
e.g., quantum error correcting codes \cite{shor} or
decoherence-free subspaces \cite{chua} are realized with
high performance, the proposed scheme seems to be
impractical.
And even if such methods are available, errors will remain
to be generated with a small rate. Alice might insist that
all errors are the residual ones and would not give him
the $R$ coins. Bob's practical solution to this problem is
that he aborts the whole protocol if the error rate is
greater than the expected residual error rate, as suggested
in the original work \cite{gold}.
Despite these difficulties,
however, it is worthwhile to have another application of
the fundamental property that nonorthogonal qubits cannot
be distinguished with certainty
\cite{levi,hole,ivan,pere,hutt,mass}.

In conclusion,
we have given another (remote)
quantum gambling scheme that makes use of
the fact that nonorthogonal states cannot be
distinguished with certainty.
In the proposed scheme, two participants Alice and Bob
can be regarded as playing a game of making guesses
on identities of quantum states that are in one of two
given nonorthogonal states:
if Bob makes a correct (an incorrect)
guess on the identity
of a quantum state that Alice has sent, he wins (loses).
It was shown that the proposed scheme is secure
against the nonentanglement attack. It could also be shown
heuristically that the scheme is secure in the case of
the entanglement attack.

\acknowledgments
This work was supported by the Korean Ministry of Science
and Technology through the Creative Research Initiatives
Program under Contract No. 99-C-CT-01-C-35.


\begin{references}
\bibitem{fnt1} We confine ourselves to this case for simplicity,
               although any two nonorthogonal qubits work.
\bibitem{fnt2} Regarding $R$ as a penalty for trial to cheat
               helps one to understand the scheme. However,
               as properly noted in Ref.\cite{gold},
               Alice's preparation of states other than the 
               specified ones is in itself gambling.
\end{references}
\end{document}